\documentclass[preprint,
pra,a4paper,showpacs,amsmath,amssymb,floatfix]{revtex4}
\usepackage{graphicx,bm}
\newcommand{\eq}[1]{Eq.~(\ref{#1})}
\newcommand{\be}{\begin{equation}}
\newcommand{\ee}{\end{equation}}
\newcommand{\ba}{\begin{eqnarray}}
\newcommand{\ea}{\end{eqnarray}}
\newcommand{\bs}{\begin{subequations}}
\newcommand{\es}{\end{subequations}}
\newcommand{\bw}{\begin{widetext}}
\newcommand{\ew}{\end{widetext}}
\newcommand{\tit}{\textit}
\newcommand{\trm}{\textrm}
\newcommand{\tbf}{\textbf}

\begin{document}

\title{Gyroscopic effects in interference of matter waves}

\author{Oleg I. Tolstikhin$^1$,
Toru Morishita$^2$, and Shinichi Watanabe$^2$}
\affiliation{$^1$Russian Research Center ``Kurchatov Institute'',
Kurchatov Square 1, Moscow 123182, Russia\\
$^2$Department of Applied Physics and Chemistry,
University of Electro-Communications,\\
1-5-1, Chofu-ga-oka, Chofu-shi, Tokyo 182-8585, Japan}

\date{\today}

\begin{abstract}
A new gyroscopic interference effect stemming from the Galilean
translational factor in the matter wave function is pointed out.
In contrast to the well-known Sagnac effect that stems from the
geometric phase and leads to a shift of interference fringes, this
effect causes slanting of the fringes. We illustrate it by
calculations for two split cigar-shaped Bose-Einstein condensates
under the conditions of a recent experiment, see Y.~Shin \tit{et
al.}, \prl \tbf{92}, 050405 (2004). Importantly, the measurement
of slanting obviates the need of a third reference cloud.
\end{abstract}
\maketitle

>From the spinning mass gyroscope to modern light and matter wave
interferometers is a spectacular evolution in man's technology for
sensing a change in orientation. From navigation of ships and
spacecrafts to studying the seismic motion of tectonic plates, the
applications of gyro keep motivating further scientific and
engineering progress in the field. It thus merits to point out in
this Communication a new gyroscopic interference effect that stems
from the Galilean translational factor in the matter wave
function.

The rotation-sensitive interference experiments have a long
history. A rotation-induced difference in optical paths for two
light beams counter-propagating around a closed contour causes a
shift of the interference fringes, which was conceived by Sagnac
as a subject of interferometry \cite{Sagnac}. This first
demonstration of the Sagnac effect was performed on a rotating
table. Soon afterward, the first measurement of the Earth's
rotation was done by Michelson and Gale \cite{MG} with a very
large stationary light interferometer operating on the same
principle. The advent of lasers ensued a rapid progress in the
optical measurements of the Sagnac effect and culminated with the
creation of miniature ring laser gyros currently in use for
navigation, see a review \cite{Scully}. The Sagnac effect for
matter waves was theoretically predicted in \cite{Page}. The
result can be obtained by changing light to matter via the
substitution $\hbar\omega=mc^2$ into the expression for the Sagnac
phase shift. The advantage of matter waves is in the possibility
of considerably down-sizing the interferometer owing to the
reduced wavelength. The effect was demonstrated experimentally
first with neutrons \cite{WSC,Atwood} and later with neutral atoms
\cite{Riehle} and charged particles (electrons) \cite{HN}. There
is also a suggestion for hybridizing light and matter waves by way
of polaritons \cite{Fleischhauer}. Highest precision and stability
to date were achieved with atom interferometers: on a rotating
table, at rates of rotation of the order of Earth's rate
\cite{Prit}, and with stationary apparatus, detecting the rotation
of Earth \cite{Kas}. It can hardly be overemphasized that
Bose-Einstein condensates (BECs) provide a new type of matter wave
sources. Their high brightness and coherence make them to matter
wave interferometry what lasers are to optical interferometry. A
single measurement of interference with BECs should suffice to
determine the rotation velocity; this paper explores BEC as a
gyroscope for this specific reason apart from a further
miniturization. The first demonstration of interference between
two BECs was reported in \cite{Ket}. Interference of BECs in
lattices was also observed \cite{vertical,lattice}. Recently, two
cigar-shaped BECs in a double-well potential were made to
interfere keeping the relative phase of atomic clouds under
control \cite{2well}. The authors conclude: ``Propagating the
separated condensates along a microfabricated wave guide prior to
phase readout would create an atom interferometer with an enclosed
area, and hence with rotation sensitivity''. Thus
rotation-sensitive BEC interferometers should become available
soon.

In all the above experiments, done and planned, rotation reveals
itself via a shift of interference fringes owing to the Sagnac
effect. In this paper we show that there is another effect that
leads to slanting of the fringes.

\begin{figure}[t]
\includegraphics[width=70mm]{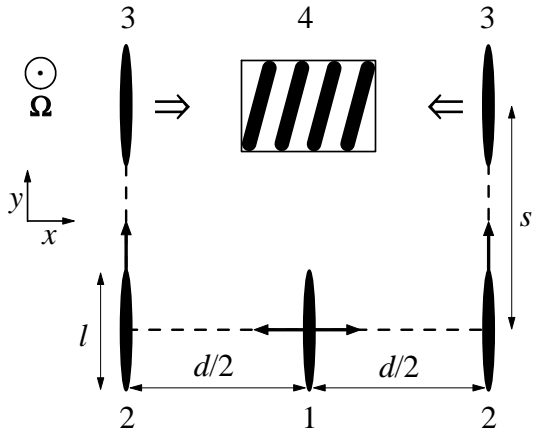}
\caption{\label{f1} General scheme of a rotation-sensitive
interference experiment with BEC. A BEC, initially in a single
cigar-shaped trap (1), is split coherently in the transverse
direction into two parts (2), which are then moved in the
longitude direction and released to expand freely (3). The
interference pattern (4) is obtained by measuring the atomic
density when the clouds begin to overlap.}
\end{figure}

As noted above, the discussion to follow deals with BEC as a
bright source of matter waves, so that we base our consideration
on the Gross-Pitaevskii (GP) equation \cite{DGPS}, but the
argumentation applies also to a dilute atomic cloud [that is
including the limit of $\beta=0$ in \eq{GP}].
The phenomenon we present here
is thus of a rather general nature, but our specific reference to
BEC as a matter wave source inevitably leads us to include
dynamical effects of the nonlinear term. The positive internal
pressure expressed by the nonlinear term is conducive to rapid
expansion of the system when the trap is switched off. The GP
equation in a rotating reference frame reads
\be \label{GP}
i\hbar\frac{\partial \psi(\tbf{r},t)}{\partial t}=
\left[\frac{\tbf{p}^2}{2m}
-\bm{\Omega} \cdot \tbf{l}
+V(\tbf{r},t)
+\beta|\psi(\tbf{r},t)|^2
\right]\psi(\tbf{r},t).
\ee
Here $\bm{\Omega}$ is the angular velocity of rotation,
$V(\tbf{r},t)$ is the trap potential, $\beta=4\pi N\hbar^2 a/m$ is
the coupling constant, where $m$, $a$, and $N$ are the atomic
mass, scattering length, and the number of atoms, respectively,
and the order parameter $\psi(\tbf{r},t)$ is normalized to unity.
We shall assume that $\bm{\Omega}$ is small and does not depend on
time. Let us consider a simple \tit{gedanken} experiment to detect
rotation, first with the aid of an approximate representation of
the experimental steps, and then Eq.~(\ref{GP}) will be
numerically solved subject to realistic experimental conditions.
The experiment proceeds in four steps, see Fig.~\ref{f1}: (1)
preparation of the initial state, which we assume to be the ground
state of a BEC in a single-well trap, (2) coherent splitting of
the BEC into two or more components by deforming the potential
into several separated traps, (3) independent transportation of
the components along some paths by varying the positions of the
traps, and (4) switching off the potential and observation of the
interference pattern when the different components begin to
overlap as a result of their free expansion. Assuming that steps 2
and 3 are done adiabatically slowly the effects of rotation on the
interference pattern can be deduced analytically. The
\tit{adiabaticity} means that the system remains in the ground
state at each moment; we will return to this point in the
discussion of numerical results. During step 3 the solution to
\eq{GP} can be decomposed as
\be \label{psi1}
\psi(\tbf{r},t)=\sum_n \psi_n(\tbf{r},t),
\ee
where the overlap of the different components $\psi_n(\tbf{r},t)$ is
geometrically marginal but physically non-negligible due
to tunneling. The position of the $n$-th component is given by
\be \label{an}
\tbf{a}_n(t)=
\frac{\left<\psi_n^*(\tbf{r},t)\tbf{r}\psi_n(\tbf{r},t)\right>_n}
{\left<\psi_n^*(\tbf{r},t)\psi_n(\tbf{r},t)\right>_n},
\ee
where $\left< \cdots \right>_n$ denotes integration over the
region occupied by the $n$-th trap. In this region
$V(\tbf{r},t)=V_n(\tbf{r}-\tbf{a}_n(t),t)$, where
$V_n(\tbf{r},t)$ defines the shape of the $n$-th trap.
If $\tbf{a}_n(t)$ and $V_n(\tbf{r},t)$ are
slowly varying functions of $t$, in the adiabatic approximation
we obtain
\be \label{psin}
\psi_n(\tbf{r},t)=\exp\left[-i\hbar^{-1}
\int^t\mu(t')\,dt'+i\gamma_n(t)\right]
\phi_n(\tbf{r};t),
\ee
where $\mu(t)$ and $\phi_n(\tbf{r};t)$ constitute
the ground state solution of the stationary equation
\be \label{sGPa}
\left[\frac{\tbf{p}^2}{2m}
-\bm{\Omega} \cdot \tbf{l}
+V_n(\tbf{r}-\tbf{a}_n(t),t)
+\beta|\phi_n(\tbf{r};t)|^2-\mu(t)
\right]\phi_n(\tbf{r};t)=0
\ee
that depends on $t$ as a parameter,
and $\gamma_n(t)$ is the geometric phase \cite{Berry},
\be \label{gam1}
\gamma_n(t)=i\int^t
\frac{\left<\phi_n^*(\tbf{r};t')
\partial\phi_n(\tbf{r};t')/\partial t'\right>_n}
{\left<\phi_n^*(\tbf{r};t')\phi_n(\tbf{r};t')\right>_n}\,dt'.
\ee
Note that the chemical potential $\mu(t)$ does not depend on $n$
and corresponds to the ground state of the whole system, which
assumes that equilibrium is attained at each instant of time via
tunneling between the components \cite{Josephson}. Our key
observation is that the solution to \eq{sGPa} can be presented in
the form
\bs \label{shift}
\ba
&&\mu(t)=\bar{\mu}(t)+O(\Omega^2), \label{shifta} \\
&&\phi_n(\tbf{r};t)=
\exp\left[i\hbar^{-1}m\tbf{v}_n(t) \cdot \tbf{r}\right]
\bar{\phi}_n(\tbf{r}-\tbf{a}_n(t);t)+O(\Omega^2), \label{shiftb}
\ea
\es
where $\bar{\mu}(t)$ and $\bar{\phi}_n(\tbf{r};t)$ constitute the
ground state solution of
\be \label{sGP}
\left[\frac{\tbf{p}^2}{2m}
-\bm{\Omega} \cdot \tbf{l}
+V_n(\tbf{r},t)
+\beta|\bar{\phi}_n(\tbf{r};t)|^2-\bar{\mu}(t)
\right]\bar{\phi}_n(\tbf{r};t)=0,
\ee
and $\tbf{v}_n(t)=\bm{\Omega} \times \tbf{a}_n(t)$.
The exponent in \eq{shiftb} is the Galilean translational
factor; it accounts for the fact that the $n$-th component
moves with the velocity $\tbf{v}_n(t)$
with respect to a non-rotating inertial frame
whose origin coincides with that of the rotating frame.
The solution of \eq{sGP} may have a nonzero phase
$O(\Omega)$ which, however, disappears
if the potential $V_n(\tbf{r},t)$ becomes axially
symmetric about the direction of $\bm{\Omega}$.
The validity of Eqs.~(\ref{shift}) can be confirmed for
the special case of a harmonic oscillator trap.
Substituting into Eqs.~(\ref{sGPa}) and (\ref{sGP})
$V_n(\tbf{r},t)=V_\text{HO}(\tbf{r})=
\frac{1}{2}\,m \tbf{r}^T\hat{\omega}^2\tbf{r}$
and $\tbf{a}_n(t)=\tbf{a}$,
where
$\hat{\omega}=\mbox{diag}\left(\omega_x,\omega_y,\omega_z\right)$
is a diagonal matrix,
$\tbf{r}$ is treated as a column, and $\tbf{r}^T$
denotes the corresponding row, one can obtain
an exact relation between the solutions
to these equations,
\bs \label{HO}
\ba
&&\mu=\bar{\mu}-\frac{m\tbf{u}^2}{2}+
V_\text{HO}(\tbf{b}-\tbf{a}), \\
&&\phi(\tbf{r})=
\exp\left[i\hbar^{-1}m\tbf{u} \cdot
\tbf{r}\right]\bar{\phi}(\tbf{r}-\tbf{b}),
\ea
\es
where $\tbf{u}=\bm{\Omega} \times \tbf{b}$ and
$\tbf{b}=
\left(\hat{\omega}^2+\bm{\Omega}\bm{\Omega}^T-
\bm{\Omega}^2\right)^{-1}
\hat{\omega}^2\tbf{a}$.
It can be easily seen that $\tbf{b}=\tbf{a}+O(\Omega^2)$,
so the result (\ref{HO}) is in agreement with
Eqs.~(\ref{shift}).
>From Eqs.~(\ref{an}), (\ref{gam1}), and (\ref{shiftb})
we find \cite{Page}
\be \label{gam2}
\gamma_n(t)
=\hbar^{-1}m
\int^{\tbf{a}_n(t)} \bm{\Omega} \times \tbf{a} \cdot d\tbf{a},
\ee
where the integral is calculated along the path traced by
the vector $\tbf{a}_n(t)$.
Summarizing, we can specify \eq{psi1} as follows
\be \label{psi2}
\psi(\tbf{r},t)=
\exp\left[-i\hbar^{-1}\int^t \bar{\mu}(t')\,dt'\right]
\sum_n \exp\left[i\gamma_n(t)
+i\hbar^{-1}m\tbf{v}_n(t) \cdot \tbf{r}\right]
\bar{\phi}_n(\tbf{r}-\tbf{a}_n(t);t).
\ee
This function provides an initial state for free expansion during
step 4. The resulting interference pattern depends on the phases
of the components in \eq{psi2}, and hence is sensitive to
rotation. A difference between the geometric phases leads to a
shift of the interference fringes; this is the Sagnac effect. The
Galilean factors introduce an additional phase difference; this
phase is inhomogeneous, which causes a deformation of the
interference pattern. There is also a third effect: the
interference pattern rotates with respect to the rotating frame,
similarly to the rotation of the plane of motion of the Foucault
pendulum. It is the last two effects that may result in slanting
of the fringes, as is shown below, and their manifestation should
be clearly distinguished from the Sagnac effect.

We illustrate this general consideration by calculations for a
two-dimensional system modeling the experiment discussed above.
Number as well as phase fluctuations in BECs are supposedly
problematic in measuring a shift of the fringes, but do not affect
their slanting. So we use the GP equation without explicit
representation of the fluctuations \cite{Josephson}.
The results reported below were obtained from \eq{GP} directly,
with no approximation or ansatz to the GP equation, by propagating
the solution in time in a finite spatial rectangle $-d \leq x \leq
d$, $-l/2 \leq y \leq l/2$, with the condition $\psi(x,y,t)=0$ on
its boundaries.  A
BEC containing
$N=10^4$ atoms $^{23}$Na ($m=3.82 \times 10^{-26}$ kg,
$a=2.75 \times 10^{-3}$ $\mu$m) is initially
in a single-well trap, see Fig.~\ref{f1}.
The trap is formed by a Gaussian potential
$V_0(x)=-U_0\exp(-\frac{1}{2}m\omega^2 x^2/U_0)$,
in the $x$ direction,
and zero boundary conditions at $y=\pm l/2$,
in the $y$ direction, with the trap depth
$U_0=h \times 5$ kHz, frequency $\omega=2\pi \times 615$ Hz,
and length $l=23$ $\mu$m.
In the first stage of the calculations, the ground
state in the trap is obtained by propagating \eq{GP}
in imaginary time.
The atomic density in this state has a cigar-like
shape with the width 6.8 $\mu$m on the level of 1\%
of the maximum value.
In the second stage,
the atomic cloud is split in the $x$ direction
into two equal cigar-shaped parts
by varying the potential according to
$V(x,t)=f(t)[V_0(x-vt)+V_0(x+vt)]$
with a constant velocity $v$
during time $\tau_x=5$ ms,
where the factor $f(t)$ is introduced to preserve trap's depth
at each moment equal to $U_0$, with $f(0)=1/2$ and
$f(\tau_x) \approx f(\infty)=1$.
The final distance between the cigars is
$d=2v\tau_x=13$ $\mu$m, and their width is 5.6 $\mu$m,
so to a very good approximation
(i) cigars do not overlap, and (ii) the zero boundary
conditions at $x=\pm d$
do not affect the solution of \eq{GP}.
Similar coherent splitting of a cigar-shaped BEC into two parts
was realized in a recent experiment \cite{2well}. The above values
of $\omega$, $U_0$, $d$, and $\tau_x$ coincide with the values of
these parameters in \cite{2well}, but in order to facilitate the
calculations the cigar length $l$, and hence the number of atoms
$N$, are taken to be equal to about one tenth of their values in
\cite{2well}. In the third stage, the cigars are transported in
the $y$ direction with the same constant velocity $v$ during
time $\tau_y$. This is done by introducing into the Hamiltonian in
\eq{GP} an additional term $-vp_y$. The transportation distance
$s=v\tau_y$ is chosen to be equal to cigar's length, $s=23$
$\mu$m, which corresponds to $\tau_y=17.7$ ms.
\begin{figure}[t]
\includegraphics[width=75mm]{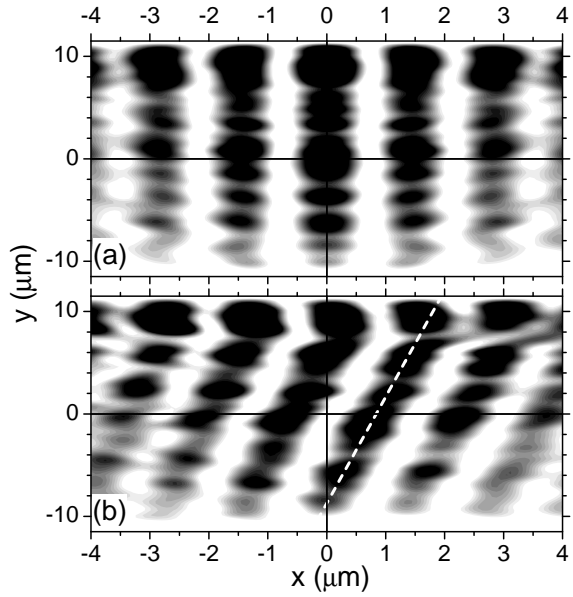}
\caption{\label{f2} An interference pattern obtained by simulation
of the experiment shown in Fig.~\protect\ref{f1} under the
conditions similar to that realized recently \protect\cite{2well}
without rotation (a) and with rotation (b). In the latter case the
interference fringes are shifted (the Sagnac effect) and slanted;
the white dashed line is a fit to the central fringe according to
Eq.~(\protect\ref{n}).}
\end{figure}
Finally, in the fourth stage, the potential is switched off and
the atomic clouds are allowed to expand freely during time
$\tau=1$ ms. The boundaries at $x=\pm d$ still do not affect the
results because this time is too short for the waves reflected
from them could reach the central region $|x| \leq 4$ $\mu$m,
where the interference pattern is observed. The calculations were
done without rotation, $\Omega=0$, and with rotation for
$\Omega=10^{-2} \times \omega$; the results are shown in
Fig.~\ref{f2}. Because of finiteness of the velocity $v$, some
oscillations of the atomic density are excited during the
splitting and transportation processes that distort the
interference pattern. Indeed, because of the cigar shape, the
density oscillation along the longitudinal axis has a low
frequency and is susceptible to nonadiabatic excitations.

\begin{figure}[t]
\includegraphics[width=75mm]{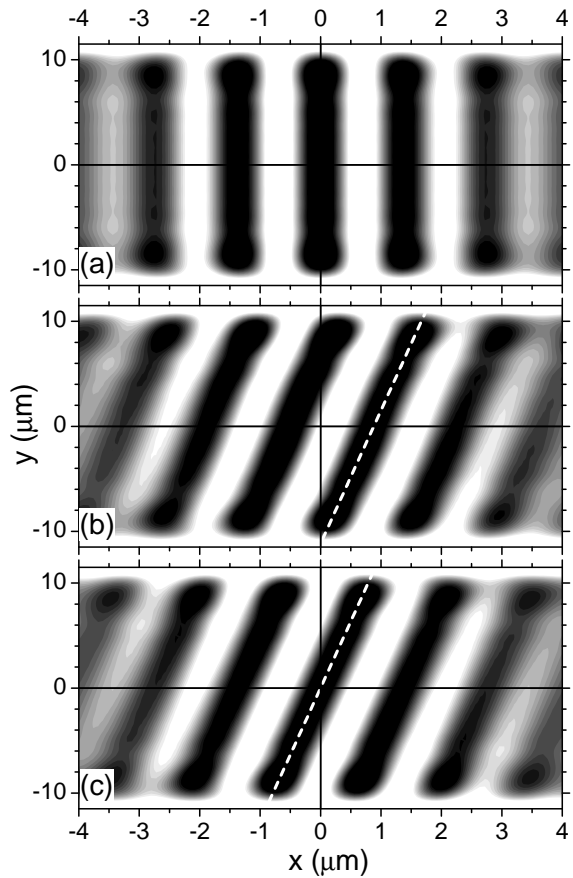}
\caption{\label{f3} (a) and (b): similar to Fig.~\protect\ref{f2},
but for 20 times slower motion during steps 2 and 3 in
Fig.~\protect\ref{f1}. (c): same as in (b), but skipping step 3;
there is no Sagnac shift in this case, but slanting remains
unchanged.}
\end{figure}

We have repeated the calculations for 20 times slower motion
($\tau_x=100$ ms, $\tau_y=354$ ms) with the same values of all the
other parameters, see Fig.~\ref{f3}. This situation is much closer
to the adiabatic regime, and the interference pattern in this case
is perfectly smooth. As can be seen from the figures, the question
of adiabaticity is of considerable importance. Elementary
excitations caused by nonadiabatic processes would quite generally
complicate the interference pattern and make the identification of
fringes difficult. Hence the experiment must be conducted slow
enough so that energy imparted to the BEC should not exceed the
excitation threshold. Anyway, in both cases one can clearly see
that in the presence of rotation the interference fringes are
shifted and slanted. We note that in real experiment the phase
shift may suffer from uncontrollable offset from run to run while
the slanting angle is known \tit{a priori} to be zero in the
absence of rotation. An interesting possibility of observing the
effect of slanting is thus assured by the very manifestation of
interference fringes.

To interpret these results, following \cite{2well} we assume that
the expansion can be described by \eq{GP} with $\beta=0$. In the
adiabatic regime, the initial condition for this equation has the
form (\ref{psi2}). Then for the atomic density
in the interference region we obtain
\be \label{n'}
n(x,y,t)=N|\psi(x,y,t)|^2
\propto 1+
\cos\left(\frac{md}{\hbar t}\,x'-\gamma_\trm{S}
-\frac{\gamma_\trm{G}}{l}\,y'\right),
\ee
where $(x,y)$ and $(x',y')$ are the coordinates of the point of
observation in the rotating and non-rotating reference frames,
respectively, and the time $t$ is measured from the moment of
release. The first term in the argument of the cosine function in
\eq{n'} corresponds to free expansion in the $x$ direction of two
point-like sources initially located at $x= \pm d/2$. The second
term comes from the geometric phases in \eq{psi2} and is given by
\be \label{gamS}
\gamma_\trm{S}=\hbar^{-1}m\Omega ds.
\ee
The third term, where
\be \label{gamG}
\gamma_\trm{G}=\hbar^{-1}m\Omega dl,
\ee
comes from the Galilean factors in \eq{psi2}.
Substituting into \eq{n'}
$x'=x\cos(\Omega t)-y\sin(\Omega t)=x-\Omega t y+O(\Omega^2)$ and
$y'=x\sin(\Omega t)+y\cos(\Omega t)=y+O(\Omega)$, we obtain
\be \label{n}
n(x,y,t) \propto 1+
\cos\left(\frac{md}{\hbar t}\,x-\gamma_\trm{S}
-\frac{2\gamma_\trm{G}}{l}\,y\right).
\ee
Thus in the presence of rotation the
interference fringes are
shifted by $\gamma_\trm{S}$, which is the Sagnac effect, and
slanted, with the phase shift between their upper and lower ends
($y=\pm l/2$) equal to $2\gamma_\trm{G}$.
These simple equations work
surprisingly well. Indeed,
according to \eq{n} the spatial period of the
fringes is $\lambda=h\tau/md=1.35$ $\mu$m,
which is very close to $\lambda^\text{calc}=1.38$ $\mu$m
obtained by fitting the numerical results.
>From Eqs.~(\ref{gamS}) and (\ref{gamG}) we have
$\gamma_\trm{S}=\gamma_\trm{G}=4.15$
(recall that in our case $s=l$).
Fitting the position of the central fringe in
Fig.~\ref{f2}(b) we obtain
$\gamma_\trm{S}^\text{calc}=3.85$ and
$\gamma_\trm{G}^\text{calc}=5.07$;
similarly from Fig.~\ref{f3}(b) we find
$\gamma_\trm{S}^\text{calc}=4.04$ and
$\gamma_\trm{G}^\text{calc}=4.13$.

To conclude, we have discussed a gyroscopic effect that stems from
the Galilean translational factor. In contrast to the Sagnac
effect that leads to a homogeneous shift of interference fringes,
the Galilean phase causes an inhomogeneous shift between the ends
of the fringes, i.e., their slanting. For the situation considered
(cigar-shaped geometry with $s \sim l$) both effects are of the
same order of magnitude. However, to produce a shift two steps are
needed, transverse separation and longitude transportation, while
for slanting the first step is sufficient, see Fig.~\ref{f3}(c).
Another important difference is that the Sagnac phase is subject
to uncertainty due to a possible asymmetry between the traps and
uncontrollable variations from run to run of the experiment which
need to be compensated for by using a third reference cloud, while
slanting is not affected by these factors.

This work was supported in part by a Grant-in-Aid for Scientific
Research (C) from the Ministry of Education, Culture, Sports,
Science and Technology, Japan, and also in part by the 21st
Century COE program on ``Coherent Optical Science''. TM was also
supported in part by a Grant-in-Aid for Young Scientist (B) from
Japan Society for the Promotion of Science and by a financial aid
from the Matsuo Foundation.


\end{document}